\documentclass[aps,prc,11pt,preprintnumbers,amsmath,amssymb,nofootinbib,groupedaddress,notitlepage,floatfix,raggedbottom]{revtex4-1}
\usepackage{amsfonts}
\usepackage{amsmath}
\usepackage{amssymb}
\usepackage{bm}
\usepackage{booktabs}
\usepackage{graphicx}
\usepackage{hyperref}
\usepackage{slashed}
\usepackage{tikz}
\usepackage{xcolor}
\usepackage{cleveref}

\newcommand{\NR}{\mathrm{NR}}
\newcommand{\lad}{\mathrm{lad}}
\newcommand{\bq}{\bm q}
\newcommand{\bk}{\bm k}
\newcommand{\bp}{\bm p}
\newcommand{\Slash}[1]{\slashed{#1}}
\newcommand{\RM}[1]{\textrm{\uppercase\expandafter{\romannumeral#1}}}

\begin{document}

\title{Binding energies in the $0^{-+}$ and $1^{--}$
fermion--antifermion channels from the ladder Bethe--Salpeter
and Schr\"odinger equations}

\author{Wei He}
\author{Hai-Qing Zhou}
\email{zhouhq@seu.edu.cn}
\affiliation{School of Physics, Southeast University,
Nanjing 211189, China}

\date{\today}

\begin{abstract}
In this work, we calculate the binding energies of an equal-mass fermion--antifermion
system in the $0^{-+}$ and $1^{--}$ channels from the ladder
Bethe--Salpeter (BS) equation with scalar, pseudoscalar, and vector exchange.
The results are compared with those of the
Schr\"odinger equation with the corresponding leading static
interaction in two regimes: pointlike
massless exchange at weak coupling and finite-mass exchange with monopole
form factors at stronger coupling.
For the BS calculation, we use two independent numerical methods based on momentum-space integral equations and coordinate-space differential equations. The differential-equation method gives estimated relative binding-energy errors below $10^{-9}$ in both regimes.
For massless vector exchange in Coulomb gauge, the
numerical results are consistent with the uncrossed-ladder perturbative
spectrum through the $O(m\alpha^5)$ terms.
The results for massless scalar exchange and vector exchange in Feynman gauge
are consistent with the common Wick--Cutkosky expansion through order
$m\alpha^3$.
For finite exchange masses, the differences from the leading static Schr\"odinger
spectra at the same masses, couplings, and form factors depend strongly
on the exchange and channel.
Scalar exchange gives nearly degenerate channels with much weaker binding than
the Schr\"odinger reference. In contrast, vector exchange with only the metric term
retained gives stronger binding than the Schr\"odinger reference in $0^{-+}$ and
weaker binding in $1^{--}$.
Notably, retaining the full Proca numerator substantially increases the
vector binding in both channels.
Pseudoscalar exchange also supports bound states and favors $1^{--}$,
although its leading static central interaction vanishes.
\end{abstract}

\maketitle

\section{Introduction}
\label{sec:introduction}

The Bethe--Salpeter (BS) equation provides a Lorentz-covariant description
of relativistic two-body bound states in quantum field theory
~\cite{BetheSalpeter1951,Nakanishi1969}.  In the weak-binding and
weak-coupling limit, a suitable ladder BS equation reduces to a
Schr\"odinger or Lippmann--Schwinger equation
~\cite{Scarf1955,Wanders1956,BhaleraoGurvitz1983,CaswellLepage1978}
and yields the same leading energy spectrum.  This formal limiting relation,
however, does not determine the accuracy of the
three-dimensional approximation at finite parameters.

In atomic physics, the Schr\"odinger equation provides an accurate description of
bound-state spectra, where relativistic and radiative
corrections can be organized systematically.  In hadronic physics, the
Schr\"odinger equation and three-dimensional Lippmann--Schwinger equations
are also widely used.  Such calculations are based on three-dimensional
reductions of covariant BS equations or on interactions whose parameters
are fixed by low-energy data. Finite exchange masses, spin-dependent
operators, and regulator dependence can be important in these systems. It is
therefore useful to quantify the differences between a three-dimensional
reduction and the corresponding BS spectrum.

The BS--Schr\"odinger comparison has been studied in equal-mass two-fermion
models by solving the integral BS equation. Dorkin et al.\ solved the ${}^1S_0$ and
${}^3S_1$--${}^3D_1$ channels in Euclidean space with scalar,
pseudoscalar, and vector exchange and monopole form factors, and compared
the results with nonrelativistic and light-front
calculations~\cite{Dorkin2008}.
Carbonell and Karmanov, as well as de Paula et al., solved the $J=0$
two-fermion channel in Minkowski space with vertex form factors
~\cite{CarbonellKarmanov2010,dePaula2016,dePaula2017}.
The numerical comparisons show substantial differences from the leading
Schr\"odinger spectrum~\cite{Dorkin2008,ManginBrinet2003}.
For massive vector exchange, the BS calculations of
Refs.~\cite{Dorkin2008,dePaula2017} retain only the metric term in the
propagator numerator, whereas related light-front calculations retain
its momentum-dependent part~\cite{ManginBrinet2003}.

As an alternative to the integral formulation, the BS equation was previously solved in differential form for special coupling combinations with massless exchange~\cite{Suttorp1978}.
In this work, we solve the ladder BS equation for an equal-mass
fermion--antifermion system using both the momentum-space integral equations and
the coordinate-space partial differential equations (PDEs) discretized by
finite elements.
The finite-element calculation gives estimated relative binding-energy
errors below $10^{-9}$. At weak coupling, we first consider
pointlike, strictly massless exchange, including scalar
exchange and massless vector exchange in Coulomb and Feynman gauges in
the $0^{-+}$ and $1^{--}$ channels. As an analytic check, we compare
these results with the corresponding perturbative expansions, including
the uncrossed Coulomb-gauge spectrum through fifth order. We then consider scalar,
pseudoscalar, and vector exchange with hadronic-scale exchange-mass
ratios and monopole vertex form factors.  For scalar and vector exchange,
we compare the BS spectra with the leading static Schr\"odinger spectra at the
same bare parameters.  For massive vector exchange, we also retain the
full Proca numerator and compare it with the metric-only prescription
at the same bare parameters in both channels.

The remainder of the article is organized as follows.  In
Sec.~\ref{sec:framework}, we present the BS equations for the $0^{-+}$ and
$1^{--}$ channels and derive the corresponding leading Schr\"odinger
equation and potentials.  Section~\ref{sec:method} describes the numerical
methods used to solve the BS equation.  Section~\ref{sec:results} presents
the strictly massless weak-coupling study and the finite-mass benchmark
results with form factors.  Section~\ref{sec:summary} summarizes the main conclusions.

\section{Formalism}

\label{sec:framework}

\subsection{Bethe--Salpeter equation for a fermion--antifermion system and the ladder interaction}
\label{subsec:BS-ladder}

For a fermion and antifermion each of mass $m$, we define the momenta in the
bound-state rest frame as
\begin{align}
  k_a\equiv\frac{P}{2}+k,\qquad
  k_b\equiv\frac{P}{2}-k,\qquad
  q_a\equiv\frac{P}{2}+q,\qquad
  q_b\equiv\frac{P}{2}-q,\qquad \nonumber\\
  P\equiv(M,\bm 0), \qquad k\equiv(k_0,\bk),\qquad q\equiv(q_0,\bq),\qquad
  k_s\equiv|\bk|,\qquad q_s\equiv|\bq|.
  \label{eq:kinematics}
\end{align}
The unamputated BS amplitude $\chi$ and the amputated vertex $\Gamma$ are
defined in the direct fermion--antifermion matrix convention as
\begin{align}
  \chi_{\alpha\beta}(k;P)
  &\equiv
  \int d^4x\,e^{ik\cdot x}
  \langle 0|T\!\left\{
  \psi_\alpha(x/2)\bar\psi_\beta(-x/2)
  \right\}|P\rangle
  \nonumber\\
  &\equiv
  \left[S(k_a)\,\Gamma(k;P)\,S(-k_b)\right]_{\alpha\beta},
  \label{eq:chi-Gamma}
\end{align}
where $|P\rangle$ is the bound state and the minus sign in $S(-k_b)$
follows the antifermion charge flow.  The
homogeneous vertex satisfies the BS equation
\begin{align}
  \Gamma(k,P)=
  \int\frac{d^4q}{(2\pi)^4}\,
  K_{\text{BS}}(k,q)
  \!\left[S(q_a)\Gamma(q,P)S(-q_b)\right].
  \label{eq:BS}
\end{align}
The bracket notation $K_{\text{BS}}(k,q)[A]$ denotes a Dirac-matrix product
of the interaction vertices, the amplitude $A$, and the propagator in a
definite order. Its explicit form is defined in Eq.~\eqref{eq:K_BS} after the interaction
vertices and boson propagators are introduced.

In the numerical calculation, we approximate the constituent propagators by
free propagators,
\begin{align}
  S(\ell)=\frac{i(\Slash \ell+m)}{\ell^2-m^2+i\epsilon},
  \label{eq:constituent-propagators}
\end{align}
and use the one-boson-exchange ladder approximation for the three
interactions
\begin{align}
  L_{\mathrm{int}}^{(S)}&\equiv g_S\bar\psi\phi_S\psi,\nonumber\\
  L_{\mathrm{int}}^{(P)}&\equiv i g_P\bar\psi\gamma_5\phi_P\psi,\nonumber\\
  L_{\mathrm{int}}^{(V)}&\equiv g_V\bar\psi\gamma^\mu V_\mu\psi,
  \label{eq:interactions}
\end{align}
with $\alpha_X\equiv g_X^2/(4\pi)$.  The corresponding interaction vertices
are
\begin{align}
  \Gamma_S^{\mathrm{int}}&=i g_S,  \nonumber\\
  \Gamma_P^{\mathrm{int}}&=-g_P\gamma_5, \nonumber\\
  \Gamma_V^{\mathrm{int},\mu}&=i g_V\gamma^\mu.
  \label{eq:Gamma-and-D}
\end{align}
The pseudoscalar interaction uses the $\gamma_5$ vertex, which is
equivalent to a pseudovector vertex only between on-shell fermions.
For scalar and pseudoscalar exchange, we use the usual propagators
\begin{align}
  D_S(\ell)& \equiv \frac{i}{\ell ^2-m_S^2+i\epsilon},  \nonumber\\
  D_P(\ell)&\equiv  \frac{i}{\ell^2-m_P^2+i\epsilon}.
\end{align}
For massless vector-boson exchange, we use the Feynman- and Coulomb-gauge
propagators as separate inputs~\cite{Vairo1996}:
\begin{align}
 \begin{aligned}
 D_{V,\mathrm{Feyn}}^{\mu\nu}(\ell)
 &\equiv
 \frac{-i g^{\mu\nu}}{\ell^2+i\epsilon},
 \\
 D_{V,\mathrm{Coul}}^{\mu\nu}(\ell)
 &\equiv
 \frac{-i}{\ell^2+i\epsilon}
 \left[
 g^{\mu\nu}
 -\frac{
 (\ell\!\cdot\! n)(\ell^\mu n^\nu+n^\mu\ell^\nu)-\ell^\mu\ell^\nu
 }{(\ell\!\cdot\! n)^2-\ell^2}
 \right],
 \end{aligned}
 \label{eq:massless-vector-propagators}
\end{align}
where $n^\mu\equiv(1,\bm 0)$.  For massive vector-boson exchange, we use the
one-parameter propagator family
\begin{align}
 D_{V,\xi_L}^{\mu\nu}(\ell)
 &\equiv
 \frac{-i}{\ell^2-m_V^2+i\epsilon}
 \left(g^{\mu\nu}-\xi_L\frac{\ell^\mu\ell^\nu}{m_V^2}\right).
 \label{eq:massive-vector-propagator-family}
\end{align}
Here $\xi_L$ multiplies the longitudinal numerator and is not a gauge
parameter, since the massive vector field has no gauge symmetry in this
model. The endpoint $\xi_L=1$ gives the full Proca propagator,
$D_{V,\mathrm{Proca}}^{\mu\nu}$, whereas $\xi_L=0$ gives the simplified
metric-only propagator, $D_{V,\mathrm{app}}^{\mu\nu}$.

Using the above interactions, the bracket notation in Eq.~\eqref{eq:BS} is
explicitly written as
\begin{align}
  K_{\text{BS}}^{S}(k,q)[A]
  &\equiv
  \Gamma_S^{\mathrm{int}}A\Gamma_S^{\mathrm{int}}D_S(k-q),\nonumber\\
  K_{\text{BS}}^{P}(k,q)[A]
  &\equiv
  \Gamma_P^{\mathrm{int}}A\Gamma_P^{\mathrm{int}}D_P(k-q),\nonumber\\
  K_{\text{BS}}^{V}(k,q)[A]
  &\equiv
  \Gamma_V^{\mathrm{int},\mu}A\Gamma_V^{\mathrm{int},\nu}D_{\mu\nu}^{V}(k-q).
  \label{eq:K_BS}
\end{align}

In this work, we retain only the exchange topology represented in
Eq.~\eqref{eq:K_BS} and consider each exchange separately in the
numerical calculations.  For a same-flavor fermion--antifermion system, the
interactions in Eq.~\eqref{eq:interactions} also generate annihilation
kernels at the same order in $g_X^2$, through the pseudoscalar in
$0^{-+}$ and through the vector in $1^{--}$.
These kernels are omitted by construction.
The calculation is therefore an exchange-only ladder model rather than the
complete $O(g_X^2)$ kernel of a same-flavor system.

For binding energies in the strictly massless, weak-coupling regime, we use
the pointlike interactions above.  For the finite exchange-mass benchmarks
and the stronger couplings considered in hadronic physics, we multiply each vertex by a
form factor depending on the momentum transfer $\ell\equiv k-q$,
\begin{align}
\Gamma_{X}^{\mathrm{int}} \rightarrow \Gamma_{X}^{\mathrm{int}}F_X(\ell^2),\qquad X\in\{S,P,V\},
\label{eq:vertex_with_FF}
\end{align}
and take the monopole form
\begin{align}
  F_X(\ell^2)
  \equiv
  \frac{m_X^2-\Lambda^2}{\ell^2-\Lambda^2+i\epsilon}.
  \label{eq:CK-FF}
\end{align}
The cutoff $\Lambda$ and the monopole shape are model inputs.

In general, the BS vertex can be expanded in a complete set of
channel-dependent covariants,
\begin{align}
  \Gamma(k,P)=\sum_b f_b(k_0,k_s)B_b(k,P).
  \label{eq:vertex-general}
\end{align}
Here $B_b(k,P)$ are Dirac structures corresponding to the quantum numbers
of the bound state, and $f_b$ are scalar coefficient functions.  We use
$(k_0,k_s)$ in place of $(k\cdot P,k^2)$ as the independent variables for
$f_b$.  The index $b$ labels the finite set of basis matrices in the chosen
channel.

To obtain equations for the coefficients, we substitute
Eq.~\eqref{eq:vertex-general} into Eq.~\eqref{eq:BS} and project onto the
basis with the linear operation
\begin{align}
  \mathcal P_a[A]
  \equiv
  \frac{1}{\Omega_J}\int d\Omega_{\hat k}\,
  \mathrm{Tr}\!\left[B_a(k,P)A\right],
  \label{eq:projection-functional}
\end{align}
where $\Omega_J=4\pi$ for both channels. In the
$1^{--}$ channel the transverse Lorentz indices are contracted with
$-g^T_{\mu\nu}\equiv-(g_{\mu\nu}-P_\mu P_\nu/M^2)$ before the angular
average.

Applying $\mathcal P_a$ to Eq.~\eqref{eq:vertex-general} gives
\begin{align}
\mathcal P_a[\Gamma(k,P)]= \sum_bN_{ab}(k_0,k_s)f_b(k_0,k_s),
\end{align}
with
\begin{align}
  N_{ab}(k_0,k_s)\equiv\mathcal P_a[B_b(k,P)].
  \label{eq:overlap-matrix}
\end{align}
The matrix $N_{ab}$ need not be diagonal, but it must be nonsingular at the
quadrature points used in the calculation.  Its explicit forms for the two
channels are given in Appendix~\ref{app:projected-kernels}.

We denote the numerators of $S(q_a)$ and $S(-q_b)$ by
\begin{align}
 Q_+(q;P)\equiv \Slash q_a+m,\qquad
 Q_-(q;P)\equiv-\Slash q_b+m,
 \label{eq:fermion-numerators}
\end{align}
and
\begin{align}
 G_0(q_0,q_s;M)\equiv
 -\frac{1}{(M/2+q_0)^2-q_s^2-m^2+i\epsilon}\,
 \frac{1}{(M/2-q_0)^2-q_s^2-m^2+i\epsilon}.
 \label{eq:two-body-denominator}
\end{align}
Each basis contribution to the BS amplitude in Eq.~\eqref{eq:chi-Gamma}
can then be written as
\begin{align}
 S(q_a)B_b(q,P)S(-q_b)
 =Q_+(q;P)B_b(q,P)Q_-(q;P)\,G_0(q_0,q_s;M).
 \label{eq:propagator-factorization}
\end{align}
Thus the scalar two-body denominator can be kept outside the projected Dirac
trace.  For exchange type $X$ we define
\begin{align}
\bar K_{ab}^{(X)}(k,q,P)
&\equiv
\mathcal P_a\!\left[
K_{\mathrm{BS}}^{X}(k,q)
\left[Q_+(q;P)B_b(q,P)Q_-(q;P)\right]
\right].
\label{eq:projected-kernel}
\end{align}
The projected equation is then
\begin{align}
 \sum_b N_{ab}(k_0,k_s)f_b(k_0,k_s)
 =\sum_X\sum_b\int\frac{d^4q}{(2\pi)^4}\,
 G_0(q_0,q_s;M)\,\bar K_{ab}^{(X)}(k,q,P)f_b(q_0,q_s).
 \label{eq:projected-BS-Minkowski}
\end{align}

For a bound state with $M<2m$, a simultaneous Wick rotation of $k_0$
and $q_0$ does not cross the explicit poles of the propagators or
monopole form factors. Assuming that the BS vertex is analytic in
the swept domain, we introduce the Euclidean momenta
\begin{align}
 k_E\equiv(k_{0E},\bk)\equiv(-ik_0,\bk),\qquad
 q_E\equiv(q_{0E},\bq)\equiv(-iq_0,\bq),
 \label{eq:wick-rotation}
\end{align}
where $q_{0E}$ and $k_{0E}$ are real. After the Wick rotation, the two-body denominator becomes
\begin{align}
 G_{0E}(q_{0E},q_s;M)\equiv G_0(iq_{0E},q_s;M)
 =-\frac{1}{(M/2+i q_{0E})^2-q_s^2-m^2}\,
 \frac{1}{(M/2-i q_{0E})^2-q_s^2-m^2}.
 \label{eq:two-body-denominator-E}
\end{align}
We then perform the spatial angular integration in
Eq.~\eqref{eq:projected-BS-Minkowski} analytically and define the
angularly reduced Euclidean kernel by
\begin{align}
 \mathcal K_{ab}^{(X)}(k_{0E},k_s;q_{0E},q_s,M)
 \equiv
 \frac{i}{(2\pi)^4}\int d\Omega_{\hat q}\,
 \bar K_{ab}^{(X)}(k,q,P).
 \label{eq:radial-kernel}
\end{align}
The Euclidean coefficient equations are therefore
\begin{align}
  \sum_bN_{ab}(i k_{0E},k_s)f_b(ik_{0E},k_s)
  &=\sum_X\sum_b\int_{-\infty}^{\infty}dq_{0E}\int_0^\infty dq_s\,
  q_s^2G_{0E}(q_{0E},q_s;M)\nonumber\\
  &\quad\times\mathcal K_{ab}^{(X)}(k_{0E},k_s;q_{0E},q_s,M)
  f_b(iq_{0E},q_s).
  \label{eq:projected-BS}
\end{align}

\subsection{BS vertex in the \texorpdfstring{$0^{-+}$}{0-+} channel}
\label{subsec:0mp-formalism}

For a $0^{-+}$ bound state we define the components transverse to the total
momentum by
\begin{align}
  k_T^\mu\equiv k^\mu-\frac{k\cdot P}{P^2}P^\mu,
  \qquad
  \gamma_T^\mu\equiv\gamma^\mu-\frac{\Slash P}{P^2}P^\mu,
  \label{eq:transverse-definitions}
\end{align}
and choose four independent Dirac structures in the $0^{-+}$ channel as
\begin{align}
  B_{1,2,3,4}^{0^-}(k)\equiv
  \left\{
  \gamma_5,\quad
  \gamma_5\frac{\Slash P}{M},\quad
  \gamma_5\frac{\Slash k_T}{M},\quad
  \gamma_5\frac{[\Slash P,\Slash k_T]}{M^2}
  \right\}.
  \label{eq:basis-0mp}
\end{align}
Then the BS vertex in the $0^{-+}$ channel can be written as
\begin{align}
  \Gamma^{0^{-+}}(k,P) =
  \sum_{a=1}^{4}f_a^{0^{-+}}(k_0,k_s)B_a^{0^-}(k,P).
  \label{eq:vertex-0mp}
\end{align}
In this work we fix both the $\text{P}$ and $\text{C}$ parities of the bound state.
Charge conjugation then constrains the relative-energy parity of the
scalar functions:
\begin{align}
f_a^{0^{-+}}(-k_0,k_s)=\sigma_a^{0^{-+}}f_a^{0^{-+}}(k_0,k_s),
\label{eq:C-parity-0mp}
\end{align}
with
\begin{align}
\sigma^{0^{-+}}\equiv(+,+,-,+).
\end{align}

\subsection{BS vertex in the \texorpdfstring{$1^{--}$}{1--} channel}
\label{subsec:1mm-formalism}

The BS vertex in the $1^{--}$ channel is transverse, $P_\mu\Gamma^\mu=0$.
We choose eight independent Dirac structures in the $1^{--}$ channel,
\begin{align}
  B_1^\mu&\equiv\gamma_T^\mu,
  &B_2^\mu&\equiv\gamma_T^\mu\Slash k_T-k_T^\mu,
  &B_3^\mu&\equiv\gamma_T^\mu\Slash P,
  &B_4^\mu&\equiv\gamma_T^\mu[\Slash k_T,\Slash P]-2k_T^\mu\Slash P,
  \nonumber\\
  B_5^\mu&\equiv k_T^\mu,
  &B_6^\mu&\equiv k_T^\mu\Slash k_T,
  &B_7^\mu&\equiv k_T^\mu\Slash P,
  &B_8^\mu&\equiv k_T^\mu[\Slash k_T,\Slash P].
  \label{eq:basis-1mm}
\end{align}
Then the BS vertex in the $1^{--}$ channel can be written as
\begin{align}
  \Gamma_\mu^{1^{--}}(k,P)
  &\equiv\sum_{a=1}^{8}f_a^{1^{--}}(k_0,k_s)B_{a\mu}(k,P).
  \label{eq:vertex-1mm}
\end{align}
In ordinary nonrelativistic notation, the complete $1^{--}$ sector is
$^3S_1\oplus{}^3D_1$.  The use of eight Dirac structures does not restrict
the state to a pure $S$ wave.

Similarly, charge conjugation constrains the relative-energy parity of the
scalar functions:
\begin{align}
f_a^{1^{--}}(-k_0,k_s)=\sigma_a^{1^{--}}f_a^{1^{--}}(k_0,k_s),
\end{align}
with
\begin{align}
  \sigma^{1^{--}}
  &\equiv(+,-,+,+,+,+,-,+).
  \label{eq:C-parity-1mm}
\end{align}

\subsection{Three-dimensional reductions and nonrelativistic approximation}

\label{subsec:NR-formalism}

In this subsection, we give a leading weak-binding reduction of
Eq.~\eqref{eq:projected-BS-Minkowski}. Near threshold, we define
\begin{align}
  M\equiv2m-E_b,\qquad E_b>0,\qquad
  \mu\equiv\frac{m}{2},\qquad
  \frac{q_s}{m}=O(v),\qquad \frac{E_b}{m}=O(v^2).
  \label{eq:NR-counting}
\end{align}

The three-dimensional nonrelativistic reduction of a BS equation is not
unique~\cite{Nakanishi1969,BlankenbeclerSugar1966,Gross1969}.
Here we adopt the leading-order static approximation
\begin{align}
  G_0(q_0,q_s;M)\approx G_{\NR}(\boldsymbol{q};E_b)\delta(q_0),\qquad
  G_{\NR}(\boldsymbol{q};E_b)
  \equiv\frac{i\pi}{2m^2}
  \frac{1}{-E_b-q_s^2/(2\mu)+i0}.
  \label{eq:NR-pole-denominator}
\end{align}
This prescription is used under the relative-energy integral with the
remaining factors evaluated at $q_0=0$, not as a pointwise identity for $G_0$.

We also expand the Dirac numerators $Q_\pm$ in $q_s/m$ and $E_b/m$ and keep their leading
terms as
\begin{align}
  Q_+(q)\approx Q_+^{\mathrm{st}}\equiv 2mP_+,\qquad
  Q_-(q)\approx Q_-^{\mathrm{st}}\equiv 2mP_-,\qquad
  P_\pm\equiv\frac{1\pm\gamma^0}{2}.
  \label{eq:NR-static-numerators}
\end{align}
These leading numerators give
\begin{align}
  Q_+(q)\Gamma Q_-(q)
  \approx4m^2P_+\Gamma P_-.
  \label{eq:NR-leading-block}
\end{align}
Thus, at leading order, only the block $P_+\Gamma P_-$ of the BS vertex contributes.

We also evaluate the external BS vertex at zero relative energy. Defining
\begin{align}
  k^{\mathrm{st}}\equiv(0,\bk),\qquad
  q^{\mathrm{st}}\equiv(0,\bq),
  \label{eq:NR-static-momenta}
\end{align}
we restrict the kernel to these static momenta,
\begin{align}
  K_{\text{BS}}(k,q)\rightarrow K_{\text{BS}}^{\mathrm{st}}(\bk,\bq)
  \equiv K_{\text{BS}}(k^{\mathrm{st}},q^{\mathrm{st}}).
  \label{eq:NR-static-slice}
\end{align}
For scalar exchange, pseudoscalar exchange, and the $g^{\mu\nu}$ part of
the vector kernel, the interaction numerators have
no additional momentum dependence,
so no further approximation is made to these numerators at this step.
For the longitudinal vector numerator, the static limit removes the
temporal part. Its spatial part has no leading central contribution after
projection, since $P_\pm\gamma^iP_\pm=0$. Thus both massive-vector
propagators give the same leading central interaction. The
static vertex equation is
then
\begin{align}
  \Gamma(k^{\mathrm{st}},P)
  \approx\sum_X\int\frac{d^3\bq}{(2\pi)^4}\,
  G_{\NR}(\bq;E_b)
  K_{\text{BS}}^{X,\mathrm{st}}(\bk,\bq)
  \!\left[4m^2P_+\Gamma(q^{\mathrm{st}},P)P_-\right].
  \label{eq:NR-static-vertex}
\end{align}
We would like to mention that the quantity $\Gamma(k^{\mathrm{st}},P)$ in
Eq.~\eqref{eq:NR-static-vertex} is the amputated vertex at zero relative
energy, rather than an equal-time Salpeter wave function obtained by
integrating the BS amplitude over $k_0$.

To isolate this leading block, we define the projection
\begin{align}
  \Pi_{+-}[A]\equiv P_+AP_-,\qquad
  \bar{\Gamma}^J\equiv\Pi_{+-}[\Gamma^J(k^{\mathrm{st}},P)],
  \label{eq:Pauli-projector}
\end{align}
and apply $\Pi_{+-}$ to both sides of Eq.~\eqref{eq:NR-static-vertex}.
Using
\begin{align}
  P_+B_1^{0^{-}}P_-&=P_+\gamma_5P_-,&
  P_+B_2^{0^{-}}P_-&=-P_+\gamma_5P_-,&
  P_+B_3^{0^{-}}P_-&=P_+B_4^{0^{-}}P_-=0,
  \label{eq:NR-projection-0mp}
\end{align}
and
\begin{align}
  P_+B_1^{i}P_-&=P_+\gamma^iP_-,&
  P_+B_3^{i}P_-&=-MP_+\gamma^iP_-,&\nonumber\\
  P_+B_a^{i}P_-&=O(k_s)\quad(a\ne1,3),
  \label{eq:NR-projection-1mm-carriers}
\end{align}
one obtains
\begin{align}
  \bar{\Gamma}^{0^{-+}}(k_s)
  &\approx g_{0^{-+}}(k_s)\,P_+\gamma_5P_-,
  \nonumber\\
  \bar{\Gamma}^{i,1^{--}}(k_s)
  &\approx g_{1^{--}}(k_s)\,P_+\gamma^iP_-+O(k_s),
  \label{eq:NR-projection-1mm}
\end{align}
with
\begin{align}
  g_{0^{-+}}(k_s)\equiv
  f_1^{0^{-+}}(0,k_s)-f_2^{0^{-+}}(0,k_s),\qquad
  g_{1^{--}}(k_s)\equiv
  f_1^{1^{--}}(0,k_s)-Mf_3^{1^{--}}(0,k_s).
  \label{eq:NR-g-coefficients}
\end{align}
The same structures appear on the right-hand side of
Eq.~\eqref{eq:NR-static-vertex}, with coefficients $g_J(q_s)$.
At this order $g_J$ depends only on the magnitude of the three-momentum.
For a matrix $A$ satisfying $A=P_+AP_-$, we use the identities
\begin{align}
  \gamma^0A\gamma^0=-A,\qquad
  P_+(\gamma^i A\gamma_i)P_-=0,\qquad
  P_+(\gamma_5 A\gamma_5)P_-=0,
  \label{eq:central-Dirac-identities}
\end{align}
where a sum over spatial $i$ is understood. These identities give attractive
central interactions for scalar exchange and the temporal component of
vector exchange, whereas pseudoscalar exchange has no $O(v^0)$ central contribution.
Thus, for
pointlike vertices, the coefficients $g_J$ obey
\begin{align}
  g_J(k_s)=
  \int\frac{d^3\bq}{(2\pi)^3}\,
  \widetilde V_{\mathrm{cent}}(|\bk-\bq|)
  \frac{g_J(q_s)}{-E_b-q_s^2/(2\mu)+i0},\qquad
  J\in\{0^{-+},1^{--}\},
  \label{eq:NR-Yukawa}
\end{align}
with
\begin{align}
  \widetilde V_{\mathrm{cent}}(|\boldsymbol\ell|)
  \equiv-\frac{g_S^2}{\boldsymbol\ell^2+m_S^2}
  -\frac{g_V^2}{\boldsymbol\ell^2+m_V^2},\qquad
  \boldsymbol\ell\equiv\bk-\bq,\qquad
  \widetilde V_P=0\quad\text{at }O(v^0).
  \label{eq:central-potential}
\end{align}

By defining the Schr\"odinger wave functions as
\begin{align}
  \psi_{0^{-+}}(k_s)
  &\equiv
  \frac{g_{0^{-+}}(k_s)}{-E_b-k_s^2/(2\mu)+i0},\nonumber\\
  \psi_{1^{--}}(k_s)
  &\equiv
  \frac{g_{1^{--}}(k_s)}{-E_b-k_s^2/(2\mu)+i0},
  \label{eq:NR-wavefunction-combinations}
\end{align}
both channels reduce to the same leading momentum-space Schr\"odinger
equation,
\begin{align}
  \left(\frac{k_s^2}{2\mu}+E_b\right)\psi_J(k_s)
  +\int\frac{d^3\bq}{(2\pi)^3}\,
  \widetilde V_{\mathrm{cent}}(|\bk-\bq|)\psi_J(q_s)=0,
  \qquad J\in\{0^{-+},1^{--}\},
  \label{eq:NR-Schrodinger}
\end{align}
Within this leading NR approximation,
Eqs.~\eqref{eq:NR-g-coefficients}
and~\eqref{eq:NR-wavefunction-combinations}
give a direct relation between the BS vertex functions and the
Schr\"odinger wave functions.
The binding energies of this leading Schr\"odinger equation, denoted by
$E_b^{\mathrm{Sch}}$, are therefore spin-degenerate,
\begin{align}
  E_b^{\mathrm{Sch}}(0^{-+})=E_b^{\mathrm{Sch}}(1^{--}).
  \label{eq:NR-degeneracy}
\end{align}

With monopole form factors, the same reduction follows after the vertex
replacement in Eq.~\eqref{eq:vertex_with_FF}; the leading central potentials
are multiplied by $F_X^2(-\boldsymbol\ell^2)$ in momentum space.
The Schr\"odinger reference spectra use the same masses, couplings, and
form factors as the corresponding BS calculations.

Equations~\eqref{eq:NR-Yukawa}--\eqref{eq:NR-degeneracy} define only the
leading $O(v^0)$ static central reference, not a complete NR Hamiltonian
through relative $O(v^2)$. At this subleading order, kinetic corrections
and momentum- and spin-dependent interactions enter. Pseudoscalar
exchange first contributes at this order, including a tensor term that
can couple the ${}^3S_1$ and ${}^3D_1$ components of $1^{--}$.
These subleading operators are omitted from our finite-mass
Schr\"odinger reference calculations.

\subsection{Weak-coupling benchmarks for strictly massless exchange}
\label{subsec:breit-perturbation}
\label{subsec:feynman-ladder-weak-coupling}
For massless vector exchange, the leading binding energy of an equal-mass
$1S$ ground state is
\begin{align}
 E_b^{\mathrm{Coulomb}}=\frac{1}{4}m\alpha_V^2.
 \label{eq:coulomb-ground-state-binding}
\end{align}
In Coulomb gauge ($V_{\mathrm C}$), the Coulomb interaction and leading
relativistic corrections are described by the exchange Breit--Pauli
Hamiltonian~\cite{Malenfant1991,GreinerReinhardt}. In the center-of-mass
frame, using $r\equiv|\bm x|$ and $\hat{\bm x}\equiv\bm x/r$, it reads
\begin{align}
  H_{\mathrm{BP}}=&~\frac{\bp^2}{m}-\frac{\alpha_V}{r}
 -\frac{\bp^4}{4m^3}
 -\frac{\alpha_V}{2m^2r}
 \left[\bp^2+
 \frac{\bm x\cdot\bigl((\bm x\cdot\bp)\bp\bigr)}{r^2}\right]
 +\frac{\pi\alpha_V}{m^2}\delta^{(3)}(\bm x)
 \nonumber\\
 &+\frac{8\pi\alpha_V}{3m^2}
 \bm s_1\cdot\bm s_2\,\delta^{(3)}(\bm x)
 +\frac{3\alpha_V}{2m^2r^3}\bm L\cdot\bm S
 +\frac{\alpha_V}{4m^2r^3}S_{12}(\hat{\bm x}).
 \label{eq:coulomb-breit-hamiltonian}
\end{align}
Here $\bm S=\bm s_1+\bm s_2$, $\bm L=\bm x\times\bp$, and
$S_{12}=4[3(\bm s_1\cdot\hat{\bm x})(\bm s_2\cdot\hat{\bm x})
-\bm s_1\cdot\bm s_2]$.
The terms beyond the leading Coulomb Hamiltonian are $O(m\alpha_V^4)$
and can be evaluated by first-order bound-state perturbation theory.  For
the $0^{-+}$, ${}^1S_0$ ground state,
$\bm s_1\cdot\bm s_2=-3/4$ and the spin--orbit and tensor expectation
values vanish.  The kinetic, orbit--orbit, Darwin, and spin--spin terms in
Eq.~\eqref{eq:coulomb-breit-hamiltonian} contribute, respectively, to the
first-order energy shift,
\begin{align}
 E_b^{\mathrm{Breit}}(0^{-+})
=E_b^{\mathrm{Coulomb}}-\frac{1}{64}m\alpha_V^4(-5-8+8-16)
=E_b^{\mathrm{Coulomb}}+\frac{21}{64}m\alpha_V^4.
 \label{eq:coulomb-breit-binding-0mp}
\end{align}

For the $1^{--}$, ${}^3S_1$ ground state,
$\bm s_1\cdot\bm s_2=+1/4$, while the spin--orbit and tensor expectation
values again vanish.  The spin--spin term is then $+16/3$ in the same
units, and hence
\begin{align}
E_{b}^{\mathrm{Breit}}(1^{--})
 =E_b^{\mathrm{Coulomb}}-\frac{1}{64}m\alpha_V^4\left(-5-8+8+\frac{16}{3}\right)
 =E_b^{\mathrm{Coulomb}}-\frac{1}{192}m\alpha_V^4.
 \label{eq:coulomb-breit-binding-1mm}
 \end{align}


We isolate the uncrossed exchange and retardation contributions from
Refs.~\cite{FultonMartin1954,Labelle1997}, subtracting the lower-order
Coulomb and Breit iterations. The resulting fifth-order ladder
references are
\begin{align}
E_b^{\lad5}(0^{-+})
&\equiv m\Biggl[
\frac{\alpha_V^2}{4}
+\frac{21\alpha_V^4}{64}
-\frac{\alpha_V^5}{8\pi}
\Biggl(6\ln\frac{1}{\alpha_V}+\frac{14}{3}-4\ln 2\Biggr)
\Biggr],
\label{eq:ladder-alpha5-0mp}\\
E_b^{\lad5}(1^{--})
&\equiv m\Biggl[
\frac{\alpha_V^2}{4}
-\frac{\alpha_V^4}{192}
-\frac{\alpha_V^5}{8\pi}
\Biggl(2\ln\frac{1}{\alpha_V}-\frac{50}{9}-\frac{4}{9}\ln 2\Biggr)
\Biggr].
\label{eq:ladder-alpha5-1mm}
\end{align}
The first two terms give the Breit binding energy; the last term is
the fifth-order uncrossed-ladder correction.

For massless vector exchange in Feynman gauge ($V_{\mathrm F}$),
the leading weak-binding numerator is momentum independent. Retaining
the covariant exchange denominator then gives the scalar ladder equation
of the Wick--Cutkosky (WC) model of two scalar constituents exchanging a
massless scalar~\cite{Wick:1954eu,Cutkosky1954}, with the coupling matched
to $\alpha_V$. Spinor-numerator and negative-energy corrections enter at
relative $O(v^2)$ in the Barbieri--Remiddi framework
~\cite{BarbieriRemiddi1978,Vairo1998}. Thus the fermion--antifermion
ground-state expansion agrees with the WC expansion through
$m\alpha_V^3$ for both $J\in\{0^{-+},1^{--}\}$.
The same agreement holds for the scalar fermion--antifermion bare ladder
through $m\alpha_S^3$.
For the WC model, Feldman et al.\ identified the logarithmic
term~\cite{Feldman1973}, while Munakata obtained the complete result at
this order, including the nonlogarithmic term~\cite{Munakata1978}.
Using this WC result as the common benchmark, we write for $V_{\mathrm F}$
in our coupling convention
\begin{align}
E_b^{\mathrm{BS}}(J,V_{\mathrm F},\alpha_V)
&= E_{b,\mathrm{asy}}^{\mathrm{BS}}(J,V_{\mathrm F},\alpha_V)
+o(m\alpha_V^3),\nonumber\\
E_{b,\mathrm{asy}}^{\mathrm{BS}}(J,V_{\mathrm F},\alpha_V)
&\equiv m\left[
 \frac{\alpha_V^2}{4}
 +\frac{\alpha_V^3}{\pi}\ln\alpha_V
 +\frac{2\ln 2-1}{2\pi}\alpha_V^3
 \right],
 \label{eq:feynman-ladder-binding-asymptotic}
\end{align}
with the corresponding ratio to the Coulomb binding energy
\begin{align}
R^{\mathrm{asy}}\equiv
\frac{E_{b,\mathrm{asy}}^{\mathrm{BS}}(J,V_{\mathrm F},\alpha_V)}
{E_{b}^{\mathrm{Coulomb}}(\alpha_V)}
 =1+\frac{4}{\pi}\alpha_V\ln\alpha_V
 +\frac{2(2\ln 2-1)}{\pi}\alpha_V.
 \label{eq:feynman-ladder-ratio-asymptotic}
\end{align}
The scalar benchmark follows by replacing $\alpha_V$ with $\alpha_S$.
Thus the leading
relative corrections are proportional to $\alpha_X\ln\alpha_X$ and
$\alpha_X$ for these two ladders, but start at $O(\alpha_V^2)$ in
Coulomb gauge.

The expansions above refer to the bare-ladder approximation. In QED, the $m\alpha_V^3\ln\alpha_V$ and $m\alpha_V^3$ terms cancel when all contributions required at this order are included. This cancellation involves an infinite family of crossed diagrams~\cite{Love1978,Feldman1980I,Feldman1980II,Vairo1996}. For massless scalar exchange, NR analyses of scalar-constituent models suggest that the leading ladder logarithm also cancels when an infinite set of crossed exchanges is included~\cite{AmgharDesplanquesTheussl2001}. For fermion--antifermion scalar exchange, we have not found a corresponding proof that both third-order terms cancel.

\subsection{Coordinate-space differential forms of the ladder BS equation }
For bound states with $0<M<2m$, the ladder BS equation for the
unamputated amplitude $\chi$ defined in Eq.~\eqref{eq:chi-Gamma}
is equivalent to a coordinate-space partial differential equation.
Such differential forms for spinor constituents have been studied
previously~\cite{Ladanyi1973,Suttorp1978}.
This reformulation retains the relative-time dependence and does not make
a three-dimensional reduction.
Using the Wick rotation in Eq.~\eqref{eq:wick-rotation} and writing
\(x_E\equiv(x_{0E},\bm x)\), we define the coordinate-space amplitude by
\begin{align}
 \Phi_E(x_E)
 \equiv
 \int\frac{d^4k_E}{(2\pi)^4}\,
 e^{ik_E\cdot x_E}\chi(ik_{0E},\bm k;P).
 \label{eq:method-2-fourier}
\end{align}
Fourier transformation of the BS equation for $\chi$ then gives
\begin{align}
 \bigl(\Slash{\mathcal D_a}-m\bigr)
 \Phi_E
 \bigl(\Slash{\mathcal D_b}+m\bigr)
 +\sum_X\mathcal V_X(x_E)[\Phi_E]
 =0,
 \label{eq:method-2-differential}
\end{align}
where
\begin{align}
 \mathcal D_a
 &\equiv\bigl(M/2+\partial_{x_{0E}},\,-i\nabla\bigr),
 \nonumber\\
 \mathcal D_b
 &\equiv\bigl(M/2-\partial_{x_{0E}},\,i\nabla\bigr).
 \label{eq:method-2-Dirac-ops}
\end{align}
Here, both differential operators act on $\Phi_E$ and the Dirac matrices multiply
it from the left and right, respectively.
The right factor follows from the inverse antifermion propagator
$S^{-1}(-k_b)=i(\Slash k_b+m)$.
The interaction \(\mathcal V_X(x_E)\) is $-i$ times the Fourier transform
of the Wick-rotated exchange kernel in Eq.~\eqref{eq:K_BS}, including the
two vertex form factors when present. For example, pointlike scalar exchange gives
\begin{align}
 \mathcal V_S[\Phi_E]
 = g_S^2\,d_{S}(\rho)\,\Phi_E,
 \qquad
 d_{S}(\rho)
 \equiv
 \frac{m_S K_1(m_S\rho)}{4\pi^2\rho},
 \label{eq:method-2-scalar-example}
\end{align}
where \(K_1\) is the modified Bessel function of the second kind and
\begin{align}
 \rho\equiv|x_E|=\sqrt{x_{0E}^2+r^2}.
 \label{eq:method-2-coords}
\end{align}
Pointlike massless vector exchange in Feynman gauge gives
\begin{align}
 \mathcal V_{V_{\mathrm F}}[\Phi_E]
 =
 -g_V^2\,d_{V_{\mathrm F}}(\rho)\,\gamma^\mu\Phi_E\gamma_\mu,
 \qquad
 d_{V_{\mathrm F}}(\rho)
 \equiv
 \frac{1}{4\pi^2\rho^2}.
 \label{eq:method-2-vector-example}
\end{align}
For pointlike massless vector exchange in Coulomb gauge, the
coordinate-space interaction is
\begin{align}
 \mathcal V_{V_{\mathrm C}}[\Phi_E]
 &=
 -\frac{\alpha_V\delta(x_{0E})}{r}\gamma^0\Phi_E\gamma^0
 +\frac{g_V^2}{4\pi^2}\Bigl[
 (\rho^{-2}-d_C)\sum_{i=1}^{3}\gamma^i\Phi_E\gamma^i
 +(3d_C-\rho^{-2})
 (\bm\gamma\cdot\hat{\bm x})
 \Phi_E(\bm\gamma\cdot\hat{\bm x})
 \Bigr],
 \label{eq:method-2-coulomb-example}
\end{align}
where
\begin{align}
 d_C(x_{0E},r)
 &\equiv
 \frac{r-|x_{0E}|\arctan(r/|x_{0E}|)}{r^3}.
 \label{eq:method-2-dC}
\end{align}
Away from the origin, $d_C$ on the
coordinate axes is fixed by the corresponding limit.
On \(x_{0E}\ge 0\), the delta function contributes half of its weight on the
full line.

When the monopole form factors in Eq.~\eqref{eq:CK-FF} are considered, the regulated
denominator $F_X^2/(\ell^2-m_X^2)$ is decomposed into partial fractions.
Its Fourier transform replaces $d_S$ in
Eq.~\eqref{eq:method-2-scalar-example} by
\begin{align}
 d_X^{\mathrm{FF}}(\rho)
 \equiv
 \frac{m_XK_1(m_X\rho)-\Lambda K_1(\Lambda\rho)}{4\pi^2\rho}
 -\frac{(\Lambda^2-m_X^2)K_0(\Lambda\rho)}{8\pi^2},
 \label{eq:method-2-dFF}
\end{align}
which is finite at $\rho=0$. The pseudoscalar kernel and the
$D_{V,\mathrm{app}}^{\mu\nu}$ vector kernel contain the same function,
multiplied by the Dirac structures of Eq.~\eqref{eq:K_BS}.
Retaining the full Proca numerator gives
\begin{align}
 \mathcal V_{V,\mathrm{Proca}}[\Phi_E]
 ={}&-g_V^2\Bigl\{
 d_{00}\gamma^0\Phi_E\gamma^0
 +d_{\mathrm{sp}}\sum_{i=1}^{3}\gamma^i\Phi_E\gamma^i
 +d_{nn}(\bm\gamma\cdot\hat{\bm x})\Phi_E(\bm\gamma\cdot\hat{\bm x})
 \nonumber\\
 &-i\,d_{0n}\bigl[\gamma^0\Phi_E(\bm\gamma\cdot\hat{\bm x})
 +(\bm\gamma\cdot\hat{\bm x})\Phi_E\gamma^0\bigr]\Bigr\},
 \label{eq:method-2-proca}
\end{align}
with
\begin{align}
 \begin{gathered}
  \begin{aligned}
   d_{00}&\equiv d-\frac{h_1+h_2x_{0E}^2}{m_V^2},
   &d_{\mathrm{sp}}&\equiv -d+\frac{h_1}{m_V^2},\\[0.5ex]
   d_{nn}&\equiv\frac{h_2r^2}{m_V^2},
   &d_{0n}&\equiv\frac{h_2x_{0E}r}{m_V^2},
  \end{aligned}\\[0.5ex]
  d\equiv d_V^{\mathrm{FF}},\qquad
  h_1\equiv\frac{d'}{\rho},\qquad
  h_2\equiv\frac{d''-d'/\rho}{\rho^2},
 \end{gathered}
 \label{eq:method-2-proca-coefficients}
\end{align}
where primes denote derivatives with respect to $\rho$. At $\rho=0$, $h_1$ has an integrable logarithmic singularity, while $h_2\rho^2$ remains finite. Omitting the $1/m_V^2$ terms recovers the interaction obtained with $D_{V,\mathrm{app}}^{\mu\nu}$.

We would like to mention that the same transformation can also be applied to the WC
model of scalar constituents~\cite{Wick:1954eu,Cutkosky1954} and its
extension to finite exchange masses.
The corresponding scalar coordinate-space amplitude $\phi_E^{\mathrm{WC}}(x_{0E},\bm x)$ satisfies
\begin{align}
 \Biggl[
 \left(-\partial_{x_{0E}}^2-\nabla^2+m^2-\frac{M^2}{4}\right)^2
 -M^2\partial_{x_{0E}}^2-g_{\mathrm{WC}}^2d_S(\rho)
 \Biggr]\phi_E^{\mathrm{WC}}(x_{0E},\bm x)=0,
 \label{eq:WC-coordinate-PDE}
\end{align}
where $g_{\mathrm{WC}}$ has mass dimension one and $d_S$ is given in
Eq.~\eqref{eq:method-2-scalar-example}. For massless exchange,
$d_S=1/(4\pi^2\rho^2)$. For a partial wave with orbital quantum
numbers $(\ell,m_\ell)$, one can write
\begin{align}
 \phi_E^{\mathrm{WC}}(x_{0E},\bm x)
 =\frac{u_\ell(x_{0E},r)}{r}
 Y_{\ell m_\ell}(\hat{\bm x}),
 \label{eq:WC-partial-wave}
\end{align}
where $Y_{\ell m_\ell}$ is a spherical harmonic and $m_\ell$ is the
magnetic quantum number. This rescaling replaces $-\nabla^2$ by
$-\partial_r^2+\ell(\ell+1)/r^2$ in the radial equation.
For an $S$ wave, $\ell=0$ and $Y_{00}=1/\sqrt{4\pi}$, so the centrifugal
term vanishes.

\section{Numerical Method}
\label{sec:method}

The numerical solution of the Schr\"odinger equation is straightforward,
so we omit its details and describe the two methods for the BS equation.

\subsection{Method for the integral equation}
\label{subsec:method-1}
In the practical numerical calculation, we multiply Eq.~\eqref{eq:projected-BS} by $N^{-1}$ and introduce
an auxiliary parameter $\eta$ to write the equation as
\begin{align}
  f_a(ik_{0E},k_s)
  ={}&\eta+\sum_X\sum_{b,c}
  [N^{-1}(ik_{0E},k_s)]_{ac}
  \int_{-\infty}^{\infty}dq_{0E}\int_0^\infty dq_s\,
  q_s^2G_{0E}(q_{0E},q_s;M)\nonumber\\
  &\times\mathcal K_{cb}^{(X)}(k_{0E},k_s;q_{0E},q_s,M)
  f_b(iq_{0E},q_s),
  \label{eq:projected-BS-with-eta}
\end{align}
with an additional condition $f_{a_*}(ik_{0E}^*,k_s^*)=1$ at a selected point where this
component is nonzero, thereby fixing the overall normalization.
The charge-conjugation parities fix the coefficient functions at negative
relative energy, allowing us to solve for $k_{0E}>0$ and reconstruct the
negative-$q_{0E}$ contribution. For each trial $E_b$, we solve for $\eta$
and the coefficient functions.
The reported binding energies satisfy
\begin{align}
 \eta(E_b^{\mathrm{BS}})=0.
 \label{eq:eta-pole-condition}
\end{align}
We also solve the homogeneous BS equation as an eigenvalue
problem~\cite{Dorkin2008} to check the binding energies obtained above,
\begin{align}
  f_a(ik_{0E},k_s)
  ={}&\lambda\sum_X\sum_{b,c}
  [N^{-1}(ik_{0E},k_s)]_{ac}
  \int_{-\infty}^{\infty}dq_{0E}\int_0^\infty dq_s\,
  q_s^2G_{0E}(q_{0E},q_s;M)\nonumber\\
  &\times\mathcal K_{cb}^{(X)}(k_{0E},k_s;q_{0E},q_s,M)
  f_b(iq_{0E},q_s),
  \label{eq:projected-BS-with-lamda}
\end{align}
where the binding energy is the root of
\begin{align}
 \lambda(E_b^{\mathrm{BS}})=1.
 \label{eq:lambda-pole-condition}
\end{align}

To solve these integral equations, we use direct quadrature in the Euclidean relative energy $q_{0E}$ and radial momentum $q_s$, without expanding the coefficient functions in a special-function basis. For massless exchange, we divide the positive radial-momentum axis into four adjacent intervals,
\begin{align}
  [0,q_{\mathrm{us}}]\cup [q_{\mathrm{us}},q_{\mathrm{soft}}]
  \cup
  [q_{\mathrm{soft}},q_{\mathrm{mid}}]
  \cup
  [q_{\mathrm{mid}},q_{\max}].
\end{align}
This piecewise construction resolves separately the ultrasoft, soft, intermediate,
and ultraviolet momentum regions. In units of $m$, we take
$q_{\mathrm{us}}=\alpha_X^2$, $q_{\mathrm{soft}}=\alpha_X$,
$q_{\mathrm{mid}}=10\alpha_X$, and $q_{\max}=10000$.

On each interval $[a_j,b_j]$, we map the Gauss--Legendre nodes
$u\in(0,1)$ to physical momenta using
\begin{align}
 p_j(u)=
 \begin{cases}
  a_j+(b_j-a_j)u, & \text{linear},\\[0.5ex]
  a_j+s_j^{(p)}\sinh\!\left[u\,\operatorname{asinh}
  \!\left(\dfrac{b_j-a_j}{s_j^{(p)}}\right)\right],
  & \text{hyperbolic sine},
 \end{cases}
 \qquad p\in\{q_s,q_{0E}\}.
 \label{eq:segmented-grid-map}
\end{align}
Both $q_s$ and positive $q_{0E}$ use the same interval boundaries, with
the linear map in the first interval and the hyperbolic-sine map in the
other three. The latter use scales
$(s_2^{(p)},s_3^{(p)},s_4^{(p)})=(\alpha_X^2,\alpha_X,10\alpha_X)$.
We take $(40,20,20,60)$ positive-$q_{0E}$ nodes and
$(40,1000,200,60)$ radial nodes for every $\alpha_X$.
The boundaries and map scales vary with $\alpha_X$, so different
couplings have the same node counts but different node coordinates.

For the finite-mass calculations, in the same units we take
$q_{\mathrm{us}}=0.01$, $q_{\mathrm{soft}}=0.1$, $q_{\mathrm{mid}}=1$,
and $q_{\max}=100$, with $(40,20,20,20)$ positive-$q_{0E}$ nodes and
$(80,100,100,140)$ radial nodes. The map scales are $s_j^{(p)}=1$
for $j=2,3,4$.
We also compare the binding energies obtained with $q_{\max}=10$,
$100$, and $1000$ and find only very small differences.

We would like to mention that massless exchange requires special treatment
of the angularly integrated kernel. After the angular integration, the
projected kernel is a sum of algebraic terms and logarithms. For a
pointlike exchange the basic logarithm is
\begin{align}
 L(k_{0E},k_s;q_{0E},q_s,m_X)
 \equiv
 \log\frac{(k_{0E}-q_{0E})^2+(k_s+q_s)^2+m_X^2}
 {(k_{0E}-q_{0E})^2+(k_s-q_s)^2+m_X^2}.
 \label{eq:angular-log-minus}
\end{align}
With the monopole form factor, the same logarithm occurs with masses
$m_X$ and $\Lambda$ and is finite for either positive mass. For $m_X=0$
and $k_s>0$, its singularity at $(q_{0E},q_s)=(k_{0E},k_s)$ is integrable.
We replace its singular part by its average over a rectangle centered
at that point, with half-widths $a=\omega_E/2$ and $b=\omega_q/2$ fixed
by the physical one-dimensional quadrature weights, including the mapping
Jacobians:
\begin{align}
 \bar{L}_0^{\mathrm{sing}}(a,b)
 =\log\frac{4k_s^2}{a^2+b^2}+3
 -\frac{a}{b}\arctan\frac{b}{a}
 -\frac{b}{a}\arctan\frac{a}{b}.
 \label{eq:massless-log-cell}
\end{align}
The smooth factors, including $q_s^2$, are evaluated at the node;
no exchange-mass regulator is introduced.

For the instantaneous Coulomb component
$D_{00}^{V,\mathrm{Coul}}=i/\boldsymbol{\ell}^2$, the angularly integrated
kernel is logarithmically singular at $q_s=k_s$ for every relative energy;
we instead subtract the source value at the target node and restore its
contribution using the analytic integral of the full radial angular
moment over the containing segment.
The spatial Coulomb tensor requires two-dimensional treatment: we use
the rectangle prescription for $0^{-+}$ and compact-window subtraction
for $1^{--}$, comparing quadrature orders 24 and 32 for the latter's
integrated transverse moments.

In the practical calculation, we assess numerical stability by varying
the quadrature grids, momentum cutoff, and integration and solver
tolerances. A detailed error analysis of this approach is not presented here.

\subsection{Method for the differential equation}
\label{subsec:method-2}

Alternatively, we solve the coordinate-space differential equation
in Eq.~\eqref{eq:method-2-differential} using the finite-element method.
In the practical numerical calculation, we apply
$\mathcal C_+[A]\equiv(\Slash{\mathcal D_a}+m)A(\Slash{\mathcal D_b}-m)$
to the differential equation. This converts its free Dirac part into
the scalar fourth-order operator $\mathcal Q_M$, where
\begin{align}
 \mathcal A_M&\equiv-\partial_{x_{0E}}^2-\nabla^2+m^2-\frac{M^2}{4},
 \nonumber\\
 \mathcal Q_M&\equiv\mathcal A_M^2-M^2\partial_{x_{0E}}^2.
 \label{eq:method-2-Q}
\end{align}
To split $\mathcal Q_M$ into second-order equations, we introduce
$\Psi_E^{(1)}\equiv\mathcal Q_M^{-1}\sum_X\mathcal V_X[\Phi_E]$ and
$\Psi_E^{(2)}\equiv\mathcal A_M\Psi_E^{(1)}$. The resulting mixed system is
\begin{align}
 \Phi_E+\mathcal C_+\bigl[\Psi_E^{(1)}\bigr]&=0,\nonumber\\
 \mathcal A_M\Psi_E^{(1)}-\Psi_E^{(2)}&=0,\nonumber\\
 \mathcal A_M\Psi_E^{(2)}-M^2\partial_{x_{0E}}^2\Psi_E^{(1)}
 -\sum_X\mathcal V_X[\Phi_E]&=0.
 \label{eq:method-2-mixed}
\end{align}
Each equation contains at most second derivatives. After angular separation,
$\Phi_E$ has radial amplitudes $w_j(x_{0E},r)/r$, with four components in
$0^{-+}$ and eight in $1^{--}$. We solve on $x_{0E}\ge0$, $r\ge0$, with regularity at
$r=0$ and decay at infinity. Charge conjugation fixes the relative-time
parity: odd components vanish at $x_{0E}=0$, while even components obey
the natural weak condition, including the Coulomb-gauge delta term.
On this domain, we discretize the equations with continuous finite elements
using degree-$p$ Lagrange polynomials at Gauss--Lobatto points on independent
time and radial partitions. Rational outer elements extend both axes to infinity.
The logarithmic singularity of the Proca kernel at the origin is integrated
using Duffy-mapped triangles.

At fixed masses and coupling, we determine $E_b$ from
Eq.~\eqref{eq:eta-pole-condition} using the same normalization and
auxiliary-source procedure as in the preceding approach.

In both regimes, we estimate the relative binding-energy error
from relative changes in $E_b$ under numerical refinement. We first check
the discretization by comparing the last three
polynomial degrees and separately subdividing each finite time and radial
element at the highest degree. An empirical safety factor of two is applied
to these degree- and element-refinement contributions. We separately vary
integration orders, internal boundaries, and the starting points of the
rational outer elements, and repeat the algebraic solve with tighter tolerances.
The total estimate combines these changes with the propagated algebraic
and root-location uncertainties.
Refinement continues until the total estimate is $10^{-9}$ or less. This estimate
is independent of analytic references and the other method, but is not
a rigorous bound.

In the following, we refer to the two approaches as Method-\RM1 and
Method-\RM2, respectively. In the practical numerical calculation,
Method-\RM2 is much more computationally efficient than Method-\RM1.
The codes for both numerical methods were optimized with assistance from ChatGPT.

\section{Numerical Results and Discussion}

\label{sec:results}

Throughout this section, we set $m=1$ and quote binding energies,
exchange masses, momenta, and cutoffs in units of $m$. Dimensionful
values are recovered by multiplying the quoted values by $m$.

\subsection{Strictly massless exchange in the weak-coupling regime}
\label{sec:NR-comparison}

We first consider pointlike vertices and strictly massless exchange,
with
$X\in\{S,V_{\mathrm C},V_{\mathrm F}\}$ in both the $0^{-+}$ and
$1^{--}$ channels.  Their leading central NR interaction is Coulombic.
We compare the BS binding with the leading Coulomb value through
\begin{align}
 R_J^X\equiv
 \frac{E_b^{\mathrm{BS}}(J,X)}
      {E_b^{\mathrm{Coulomb}}},
 \label{eq:direct-BS-Yukawa-ratio}
\end{align}
where $E_b^{\mathrm{Coulomb}}$ is given by
Eq.~\eqref{eq:coulomb-ground-state-binding} with the corresponding
coupling $\alpha_X$.

\begin{figure*}[htbp]
  \centering
  \includegraphics[width=0.68\textwidth]
  {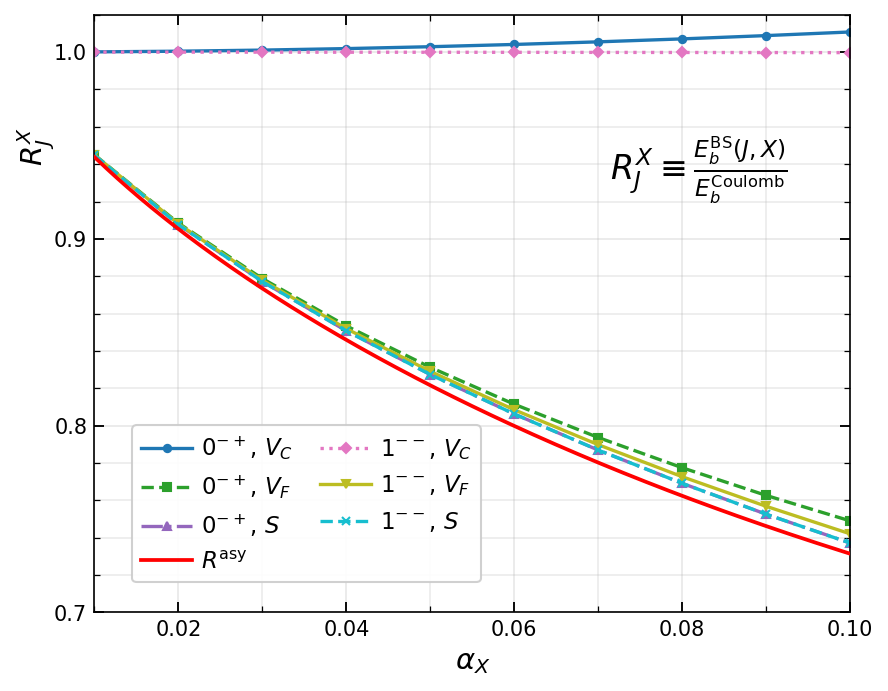}
  \caption{Binding-energy ratios $R_J^X$ for strictly massless scalar
  and vector exchange from Method-\RM2. The red curve gives
  $R^{\mathrm{asy}}$ in
  Eq.~\eqref{eq:feynman-ladder-ratio-asymptotic}.}
  \label{fig:light-bs-yuk}
\end{figure*}

Fig.~\ref{fig:light-bs-yuk} shows $R_J^X$ as a function of
$\alpha_X$. The red curve is the analytic weak-coupling ratio
$R^{\mathrm{asy}}$ in Eq.~\eqref{eq:feynman-ladder-ratio-asymptotic}.  The figure shows that the Coulomb-gauge ratios remain close to unity,
whereas the scalar and Feynman-gauge vector ratios deviate substantially
from unity and lie close to $R^{\mathrm{asy}}$.
Thus the common leading Coulomb reduction does not imply the same
subleading behavior. In the following, we discuss the Coulomb-gauge results
before turning to Feynman-gauge vector and scalar exchange.

\subsubsection{Coulomb Gauge}
Tab.~\ref{tab:coulomb-breit-comparison} compares the BS binding energies from
Method-\RM1 and Method-\RM2 with the Coulomb, Breit, and uncrossed-ladder fifth-order
references. We define the relative differences from the latter two by
\begin{align*}
\delta_{\mathrm{Breit}}^{A}
&\equiv\frac{E_b^{\mathrm{BS},A}-E_b^{\mathrm{Breit}}}
{E_b^{\mathrm{Breit}}},\qquad
\delta_{\lad5}^{A}
\equiv\frac{E_b^{\mathrm{BS},A}-E_b^{\lad5}}{E_b^{\lad5}},
\qquad A\in\{\RM1,\RM2\}.
\end{align*}

\begin{table*}[htbp]
  \caption{BS binding energies from Method-\RM1 and Method-\RM2
  for strictly massless vector exchange in Coulomb gauge,
  compared with the Coulomb, Breit, and fifth-order ladder spectra.}
  \label{tab:coulomb-breit-comparison}
  \begin{ruledtabular}
  \begin{tabular}{@{\hspace{0.35em}}l@{\hspace{0.55em}}||@{\hspace{0.15em}}c@{\hspace{0.15em}}c@{\hspace{1.30em}}||@{\hspace{-0.35em}}c@{\hspace{0.15em}}c@{\hspace{0.25em}}}
  & \multicolumn{2}{c@{\hspace{1.30em}}||@{\hspace{-0.35em}}}{$\alpha_V=0.01$}
  & \multicolumn{2}{c}{$\alpha_V=0.1$}\\
  & $0^{-+}$ & $1^{--}$ & $0^{-+}$ & $1^{--}$\\
  \colrule
  $E_b^{\mathrm{BS},\RM1}\times10^{5}$
  & $2.500316630$ & $2.499993604$ & $252.6982474$ & $249.9323868$\\
  $E_b^{\mathrm{BS},\RM2}\times10^{5}$
  & $2.500316465$ & $2.499993343$ & $252.6983257$ & $249.9325040$\\
  \colrule
  $E_b^{\mathrm{Coulomb}}\times10^{5}$
  & $2.500000000$ & $2.500000000$ & $250.0000000$ & $250.0000000$\\
  $E_b^{\mathrm{Breit}}\times10^{5}$
  & $2.500328125$ & $2.499994792$ & $253.2812500$ & $249.9479167$\\
  $E_b^{\lad5}\times10^{5}$
  & $2.500316377$ & $2.499993460$ & $252.6561853$ & $249.9979888$\\
  \colrule
  $\delta_{\mathrm{Breit}}^{\RM1}$
  & $-4.60\times10^{-6}$ & $-4.75\times10^{-7}$ & $-2.30\times10^{-3}$ & $-6.21\times10^{-5}$\\
  $\delta_{\mathrm{Breit}}^{\RM2}$
  & $-4.66\times10^{-6}$ & $-5.79\times10^{-7}$ & $-2.30\times10^{-3}$ & $-6.17\times10^{-5}$\\
  \colrule
  $\delta_{\lad5}^{\RM1}$
  & $1.01\times10^{-7}$ & $5.75\times10^{-8}$ & $1.66\times10^{-4}$ & $-2.62\times10^{-4}$\\
  $\delta_{\lad5}^{\RM2}$
  & $3.49\times10^{-8}$ & $-4.68\times10^{-8}$ & $1.67\times10^{-4}$ & $-2.62\times10^{-4}$\\
  \end{tabular}
  \end{ruledtabular}
\end{table*}

At $\alpha_V=0.01$, both channels lie close to the Coulomb spectrum, and
including the Breit and fifth-order terms successively improves agreement
with the BS results. The tabulated $|\delta_{\lad5}^{\RM2}|$ values are
$(3.49, 4.68)\times10^{-8}$ for $0^{-+}$ and $1^{--}$, respectively.
The corresponding numerical relative error estimates from Method-\RM2
are $(8.0, 7.9)\times10^{-10}$, well below these differences.
The differences from $E_b^{\lad5}$ are compatible with an $O(\alpha_V^4)$
relative remainder,
corresponding to $O(\alpha_V^6)$ in the binding energy, apart from
logarithms.
As a numerical cross-check, Method-\RM1 and Method-\RM2
differ by $(6.6,10)\times10^{-8}$ in the two channels, relative to the
Method-\RM2 binding energies.

At $\alpha_V=0.1$, the tabulated $|\delta_{\lad5}^{\RM2}|$ values are
$(1.67, 2.62)\times10^{-4}$ for $0^{-+}$ and
$1^{--}$, respectively. The corresponding numerical relative error
estimates are $(5.2, 3.4)\times10^{-10}$, much smaller
than the differences from either analytic reference.
The differences from $E_b^{\lad5}$ therefore indicate contributions
beyond fifth order rather than the estimated numerical uncertainty.
The $1^{--}$ BS result is nevertheless closer to Breit than to
$E_b^{\lad5}$. In Eq.~\eqref{eq:ladder-alpha5-1mm}, the logarithmic
and constant terms give a negative fifth-order correction at
$\alpha_V=0.01$ but a positive one at $\alpha_V=0.1$.
This positive correction raises $E_b^{\lad5}$ above Breit, whereas
the BS result lies below Breit, so the remainder beyond fifth order
has the opposite sign.

\subsubsection{Feynman-Gauge Vector and Scalar Exchange}
Tabs.~\ref{tab:feynman-asymptotic} and \ref{tab:scalar-asymptotic}
compare the Feynman-gauge vector and scalar BS binding energies with the
common asymptotic expression in Eq.~\eqref{eq:feynman-ladder-binding-asymptotic},
using $\alpha_V\to\alpha_S$ for scalar exchange. For both exchanges, we define
\begin{align}
\delta_{\mathrm{asy}}^{A}
&\equiv\frac{E_b^{\mathrm{BS},A}-E_{b,\mathrm{asy}}^{\mathrm{BS}}}
{E_{b,\mathrm{asy}}^{\mathrm{BS}}},\qquad A\in\{\RM1,\RM2\}.
\label{eq:rel-diff-asy}
\end{align}

\begin{table*}[htbp]
  \caption{BS binding energies from Method-\RM1 and Method-\RM2
  for strictly massless vector exchange in Feynman gauge,
  compared with $E_{b,\mathrm{asy}}^{\mathrm{BS}}$ in
  Eq.~\eqref{eq:feynman-ladder-binding-asymptotic}.}
  \label{tab:feynman-asymptotic}
  \begin{ruledtabular}
  \begin{tabular}{@{\hspace{0.35em}}l@{\hspace{0.55em}}||@{\hspace{0.15em}}c@{\hspace{0.15em}}c@{\hspace{1.30em}}||@{\hspace{-0.35em}}c@{\hspace{0.15em}}c@{\hspace{0.25em}}}
  & \multicolumn{2}{c@{\hspace{1.30em}}||@{\hspace{-0.35em}}}{$\alpha_V=0.01$}
  & \multicolumn{2}{c}{$\alpha_V=0.1$}\\
  & $0^{-+}$ & $1^{--}$ & $0^{-+}$ & $1^{--}$\\
  \colrule
  $E_b^{\mathrm{BS},\RM1}\times10^{5}$
  & $2.362871332$ & $2.362579410$ & $187.2506985$ & $185.5176452$\\
  $E_b^{\mathrm{BS},\RM2}\times10^{5}$
  & $2.362461377$ & $2.362169518$ & $187.2501098$ & $185.5170892$\\
  $E_{b,\mathrm{asy}}^{\mathrm{BS}}\times10^{5}$
  & $2.359560946$ & $2.359560946$ & $182.8545058$ & $182.8545058$\\
  \colrule
  $\delta_{\mathrm{asy}}^{\RM1}$
  & $1.40\times10^{-3}$ & $1.28\times10^{-3}$ & $2.40\times10^{-2}$ & $1.46\times10^{-2}$\\
  $\delta_{\mathrm{asy}}^{\RM2}$
  & $1.23\times10^{-3}$ & $1.11\times10^{-3}$ & $2.40\times10^{-2}$ & $1.46\times10^{-2}$\\
  \end{tabular}
  \end{ruledtabular}
\end{table*}

\begin{table*}[htbp]
  \caption{BS binding energies from Method-\RM1 and Method-\RM2
  for strictly massless scalar exchange, compared with
  $E_{b,\mathrm{asy}}^{\mathrm{BS}}$ in
  Eq.~\eqref{eq:feynman-ladder-binding-asymptotic},
  with $\alpha_V\to\alpha_S$.}
  \label{tab:scalar-asymptotic}
  \begin{ruledtabular}
  \begin{tabular}{@{\hspace{0.35em}}l@{\hspace{0.55em}}||@{\hspace{0.15em}}c@{\hspace{0.15em}}c@{\hspace{1.30em}}||@{\hspace{-0.35em}}c@{\hspace{0.15em}}c@{\hspace{0.25em}}}
  & \multicolumn{2}{c@{\hspace{1.30em}}||@{\hspace{-0.35em}}}{$\alpha_S=0.01$}
  & \multicolumn{2}{c}{$\alpha_S=0.1$}\\
  & $0^{-+}$ & $1^{--}$ & $0^{-+}$ & $1^{--}$\\
  \colrule
  $E_b^{\mathrm{BS},\RM1}\times10^{5}$
  & $2.362333226$ & $2.362333220$ & $184.2843785$ & $184.2831981$\\
  $E_b^{\mathrm{BS},\RM2}\times10^{5}$
  & $2.361923452$ & $2.361923447$ & $184.2838436$ & $184.2826632$\\
  $E_{b,\mathrm{asy}}^{\mathrm{BS}}\times10^{5}$
  & $2.359560946$ & $2.359560946$ & $182.8545058$ & $182.8545058$\\
  \colrule
  $\delta_{\mathrm{asy}}^{\RM1}$
  & $1.17\times10^{-3}$ & $1.17\times10^{-3}$ & $7.82\times10^{-3}$ & $7.81\times10^{-3}$\\
  $\delta_{\mathrm{asy}}^{\RM2}$
  & $1.00\times10^{-3}$ & $1.00\times10^{-3}$ & $7.82\times10^{-3}$ & $7.81\times10^{-3}$\\
  \end{tabular}
  \end{ruledtabular}
\end{table*}

At $\alpha_X=0.01$, the binding energies for both exchanges are about
$94.5\%$ of the Coulomb value, with relative differences from the common
asymptotic reference of order $10^{-3}$. For $0^{-+}$ and $1^{--}$,
respectively, the Method-\RM2 relative numerical error estimates are
$(7.9,7.7)\times10^{-10}$ for $V_{\mathrm F}$ and about
$7.7\times10^{-10}$ for scalar exchange in both channels, well below these differences.
Relative to the Method-\RM2 binding energies, the two methods differ by about
$1.7\times10^{-4}$ for both exchanges in both channels, much more than
for Coulomb-gauge exchange. These differences indicate lower accuracy of
Method-\RM1 on the present grid and limit more precise tests of
higher-order contributions with this approach.

At $\alpha_X=0.1$, the tabulated relative differences from the asymptotic
reference increase to $(2.40,1.46)\%$ for $V_{\mathrm F}$ and
$(0.782,0.781)\%$ for scalar exchange, in the same channel order.
The corresponding Method-\RM2 relative error estimates are
$(8.8,1.2)\times10^{-10}$ for $V_{\mathrm F}$ and
$(8.0,7.2)\times10^{-11}$ for scalar exchange, still much smaller than
the reference differences. These deviations therefore indicate contributions
beyond third order rather than the estimated numerical uncertainty.
The tabulated results give differences between the two methods of about
$3\times10^{-6}$ relative for both exchanges, so Method-\RM1 also
resolves these overall deviations.

The common third-order expression is independent of the channel, but the
numerical spectra show a clear splitting for $V_{\mathrm F}$ and nearly
degenerate scalar channels. The much smaller scalar splitting is
consistent with the absence of a spin--spin term in the $O(v^2)$
scalar-exchange interaction.

\subsection{Results with hadronic-scale exchange masses and form factors}
\label{sec:finite-mass-benchmark}

We now compare the ladder BS and leading static Schr\"odinger spectra for finite
exchange masses and stronger couplings, with monopole form factors.
We use the dimensionless parameters
\begin{align}
 m_V=0.77,\qquad m_S=0.50,\qquad m_P=0.14,\qquad \Lambda=2.00.
 \label{eq:benchmark-masses}
\end{align}
These values set hadronic-scale interaction ranges. The couplings and
vertices are not fitted to a specific hadronic system, and the parameters
do not define a complete $\sigma$--$\pi$--$\rho$ model. The comparisons
refer to the specified regularized exchange-only ladder kernels.
The cutoff $\Lambda$ is kept fixed here, although earlier calculations
find strong form-factor sensitivity~\cite{Dorkin2008}.

\begin{figure*}[htbp]
  \centering
  \includegraphics[width=1.0\textwidth]
  {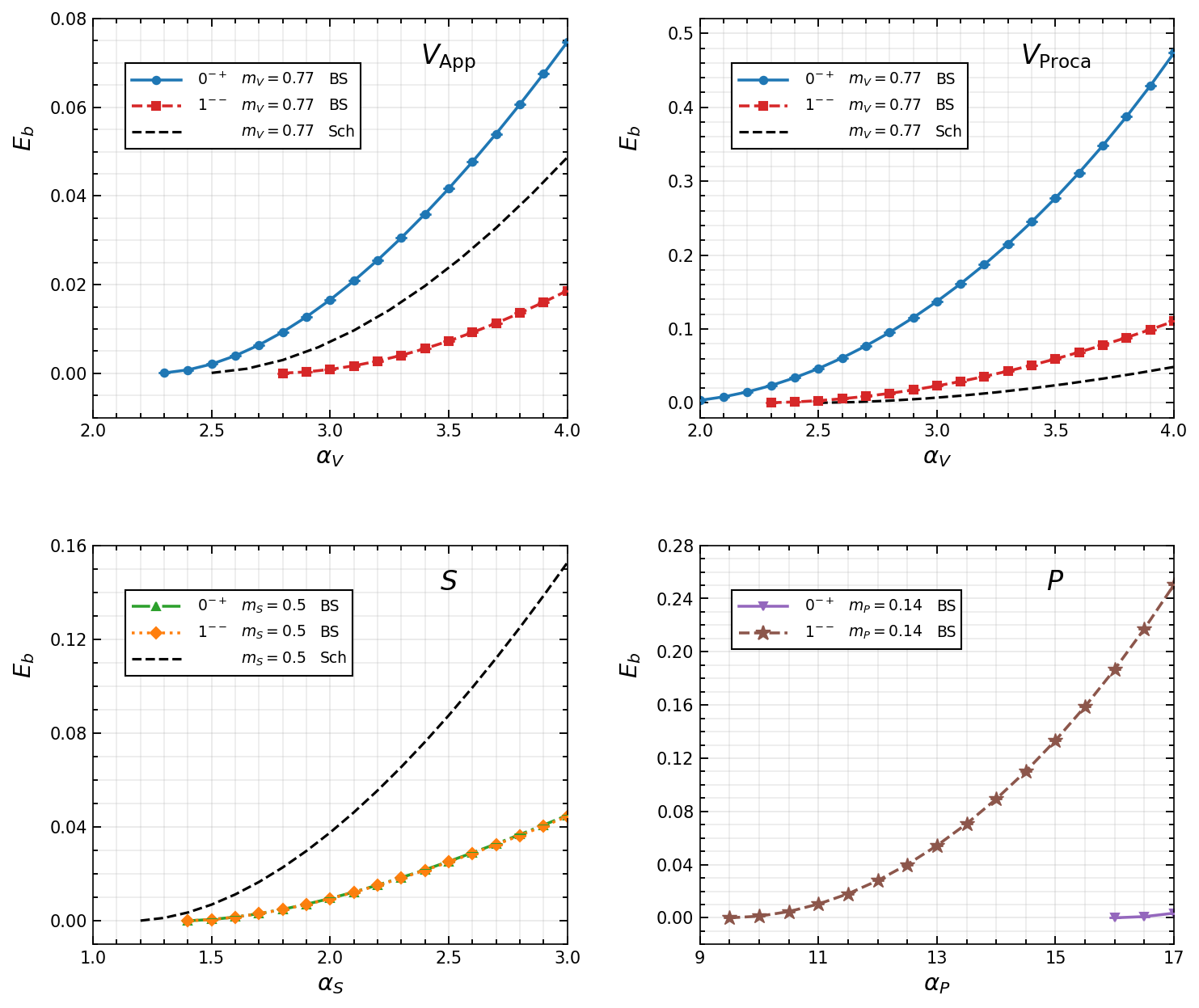}
  \caption{Binding energies from Method-\RM2 with monopole form
  factors and the parameters in Eq.~\eqref{eq:benchmark-masses}.
  The vector panels use $\xi_L=0$ and $1$ for $V_{\mathrm{App}}$
  and $V_{\mathrm{Proca}}$, respectively. Dashed black curves
  give the leading static Schr\"odinger reference.}
  \label{fig:withFF-abs}
\end{figure*}

Fig.~\ref{fig:withFF-abs} compares the $0^{-+}$ and $1^{--}$ binding
energies for both vector propagators, scalar exchange, and pseudoscalar
exchange. The plotted coupling ranges are
$2.0\leq\alpha_V\leq4.0$, $1.0\leq\alpha_S\leq3.0$, and
$9.0\leq\alpha_P\leq17.0$.

For vector exchange with $m_V=0.77$, the $0^{-+}$ state is substantially
more deeply bound than $1^{--}$ where both are resolved.
Retaining the full Proca numerator ($\xi_L=1$)
substantially strengthens the binding in both channels without changing
this ordering. The Proca $0^{-+}$ state is already bound at the lower
scan limit $\alpha_V=2.0$.

Unlike vector exchange, scalar exchange with $m_S=0.50$ gives nearly
degenerate binding energies over the resolved range. As an independent
scalar check, interpolation of the $0^{-+}$ scan in
Fig.~\ref{fig:withFF-abs} gives couplings within $2.4\times10^{-5}$ relative
of the Minkowski-space results for the same kernel and form factor
at $E_b=0.01$, $0.02$, and $0.04$~\cite{dePaula2016}.
The scalar results are also consistent with earlier Minkowski-space and
Euclidean calculations~\cite{CarbonellKarmanov2010,Dorkin2008}.

Pseudoscalar exchange, by contrast, gives a large splitting and favors
$1^{--}$. The pseudoscalar ladder binds despite the absence of an
$O(v^0)$ central attraction, as also found in earlier strong-coupling calculations with
different parameters or channels~\cite{dePaula2016}.
The $\gamma_5$ vertex directly couples positive- and
negative-energy components, without the velocity suppression of
positive-energy matrix elements. We do not solve the first nonzero
spin- and momentum-dependent NR interaction, so these results do not
rule out binding in a systematically matched NR theory.

For a quantitative comparison with the Schr\"odinger spectra, we
define the scalar and vector binding-energy ratios
\begin{align}
 R_{J,\mathrm{Sch}}^X
 \equiv
 \frac{E_b^{\mathrm{BS}}(J,X)}{E_b^{\mathrm{Sch}}(X)},
 \qquad X\in\{S,V\}.
 \label{eq:finite-BS-Yukawa-ratio}
\end{align}
Departures of these ratios from unity measure the relative BS--Schr\"odinger
differences. For vector exchange, the dependence of
$E_b^{\mathrm{BS}}$ on the specified $\xi_L$ is suppressed in the notation.

\begin{table*}[htbp]
  \caption{BS binding energies from Method-\RM2 and their ratios
  to the leading static Schr\"odinger reference, with the parameters in
  Eq.~\eqref{eq:benchmark-masses}. Estimated relative BS numerical
  errors are below $10^{-9}$. A dash denotes no resolved BS state
  or no leading static reference.}
  \label{tab:finite-mass-representative}
  \begin{ruledtabular}
  \begin{tabular}{ccccccc}
  Exchange & $\alpha_X$ & $E_b^{\mathrm{BS}}(0^{-+})$
  & $E_b^{\mathrm{BS}}(1^{--})$ & $E_b^{\mathrm{Sch}}$
  & $R_{0^{-+},\mathrm{Sch}}^X$ & $R_{1^{--},\mathrm{Sch}}^X$ \\
  \colrule
  $V(\xi_L=0)$ & 3.0 & 0.01658173989 & 0.0008790421739 & 0.007045
  & 2.354 & 0.125 \\
  $V(\xi_L=0)$ & 4.0 & 0.07466602793 & 0.01862681590 & 0.048661
  & 1.534 & 0.383 \\
  $V(\xi_L=1)$ & 3.0 & 0.1372887348 & 0.02293238699 & 0.007045
  & 19.49 & 3.255 \\
  $V(\xi_L=1)$ & 4.0 & 0.4735355418 & 0.1104177938 & 0.048661
  & 9.731 & 2.269 \\
  \colrule
  $S$ & 2.0 & 0.009594655371 & 0.009491387438 & 0.037669
  & 0.255 & 0.252 \\
  $S$ & 3.0 & 0.04499748347 & 0.04449345156 & 0.152545
  & 0.295 & 0.292 \\
  \colrule
  $P$ & 10.0 & -- & 0.001246041077 & -- & -- & -- \\
  $P$ & 17.0 & 0.003500138025 & 0.2503494340 & -- & -- & -- \\
  \end{tabular}
  \end{ruledtabular}
\end{table*}

Tab.~\ref{tab:finite-mass-representative} shows that the BS--Schr\"odinger
differences depend strongly on the exchange and channel.
The nearly degenerate scalar states have ratios near $0.25$--$0.30$,
whereas vector exchange with $\xi_L=0$ gives stronger binding than the
Schr\"odinger reference in $0^{-+}$ and weaker binding in $1^{--}$.
With the full Proca propagator, both vector channels exceed the reference.
The two vector prescriptions have the same leading static
interaction, but at $\alpha_V=4$ changing $\xi_L$ from $0$ to $1$
increases the tabulated binding energies by factors of $6.34$ and $5.93$
in the $0^{-+}$ and $1^{--}$ channels, respectively. This difference is
therefore not captured by the leading static reduction.
The weaker binding in the massless WC weak-coupling expansion,
Eq.~\eqref{eq:feynman-ladder-binding-asymptotic}, does not hold generally
for these finite-mass fermion--antifermion ladders.

The above results show that even small binding energies do not
necessarily imply an accurate leading static approximation, as
illustrated by the vector $1^{--}$ entry at $\alpha_V=3$ and $\xi_L=0$.
Physically, this can be understood by comparing the kinetic and
binding energies in the corresponding Schr\"odinger equation.
For a Coulomb ground state, the mean kinetic
energy $\langle T\rangle$ equals $E_b$ and
$\langle\bm p^2\rangle/m^2=\alpha_X^2/4$, so a small
$E_b/m$ implies small relative momenta.  For the scalar and vector reference states in
Tab.~\ref{tab:finite-mass-representative}, numerical solutions of the
Schr\"odinger equation with the regularized Yukawa interactions give
$\langle\bm p^2\rangle/m^2=0.066$--$0.277$ and
$\langle T\rangle/E_b^{\mathrm{Sch}}=1.8$--$9.3$.
Thus a small binding energy can arise from a cancellation between
larger kinetic and potential energies and need not imply small
relative momenta.

In the above calculations, we have not included crossed-kernel
contributions. They give substantial attractive corrections in
scalar-constituent models with finite exchange masses
~\cite{Theussl:1999xq,NieuwenhuisTjon1996,CarbonellKarmanov2006}.  Their effect on the BS--Schr\"odinger differences in the present
fermion--antifermion system requires further investigation.

\section{Summary}
\label{sec:summary}

We studied the equal-mass fermion--antifermion ladder BS equation
in the $0^{-+}$ and $1^{--}$ channels in two regimes: pointlike,
strictly massless exchange at weak coupling and finite-mass exchange
with monopole form factors at stronger coupling. The binding energies
were calculated using two numerical methods based on the momentum-space
integral and coordinate-space differential forms of the equation.
The finite-element solution of the latter gives estimated relative
binding-energy errors below $10^{-9}$ in both regimes. Comparison of the
two methods in the weak-coupling regime provides a numerical cross-check
and helps assess errors in the integral-equation results, offering a reference for similar
integral-equation BS calculations.

The strictly massless weak-coupling results are consistent with the
Coulomb-gauge vector ladder expansion through fifth order and the
third-order expansions for scalar and Feynman-gauge vector exchange.
The remaining differences in the differential-equation results
exceed the estimated numerical errors and suggest higher-order
contributions.

For finite exchange masses with monopole form factors at stronger
coupling, the scalar and vector BS spectra differ substantially from
their leading static Schr\"odinger references, even for small binding
energies. Scalar exchange gives nearly degenerate channels, whereas
vector and pseudoscalar exchange produce strong splitting between the
two spin channels. Although the two vector prescriptions have the same
leading static interaction, retaining the full Proca numerator gives
much stronger binding than the metric-only $V_{\mathrm{App}}$
approximation at the same bare parameters.
Crossed-ladder contributions may modify these finite-mass results,
and we have not evaluated their effects.
\appendix
\section{Projected overlap and angularly reduced kernels}
\label{app:projected-kernels}

\subsection{Expressions for $N_{ab}$}

In the rest frame we write
\begin{align}
 \kappa\equiv k_s^2=-k_T^2.
 \label{eq:app-kappa}
\end{align}
For the basis in Eq.~\eqref{eq:basis-0mp}, the angular average gives
\begin{align}
 N^{0^{-+}}
 =\operatorname{diag}\!\left(
 4,-4,\frac{4\kappa}{M^2},\frac{16\kappa}{M^2}
 \right).
 \label{eq:app-N-0mp-explicit}
\end{align}
For the $1^{--}$ basis in Eq.~\eqref{eq:basis-1mm}, the transverse
contraction is
\begin{align}
 N_{ab}^{1^{--}}(k)
 =\left(-g^T_{\mu\nu}\right)
 \mathrm{Tr}\!\left[B_a^\mu(k)B_b^\nu(k)\right],
 \label{eq:app-N-1mm-def}
\end{align}
and in the rest frame
\begin{align}
N^{1^{--}}=
\begin{pmatrix}
-12&0&0&0&0&4\kappa&0&0\\
0&-8\kappa&0&0&0&0&0&0\\
0&0&12M^2&0&0&0&0&-8\kappa M^2\\
0&0&0&-32\kappa M^2&0&0&0&0\\
0&0&0&0&4\kappa&0&0&0\\
4\kappa&0&0&0&0&-4\kappa^2&0&0\\
0&0&0&0&0&0&4\kappa M^2&0\\
0&0&-8\kappa M^2&0&0&0&0&16\kappa^2M^2
\end{pmatrix}.
\label{eq:app-N-1mm-explicit}
\end{align}
The matrices are nonsingular for $k_s>0$, including at $k_{0E}=0$.
Odd coefficient functions vanish at $k_{0E}=0$ by charge conjugation.

\subsection{Structure of the angularly reduced kernels}

We give the structure of the angularly reduced kernels defined in
Eqs.~\eqref{eq:projected-kernel} and~\eqref{eq:radial-kernel}, omitting
the lengthy coefficient expressions. The squared Euclidean momentum
transfer is
\begin{align}
 \ell_E^2(z)=(k_{0E}-q_{0E})^2+k_s^2+q_s^2-2k_s q_s z,
 \qquad z\equiv\hat\bk\!\cdot\!\hat\bq .
 \label{eq:app-ellE}
\end{align}
Representative angular integrals for pointlike scalar, pseudoscalar,
and covariant vector exchange have the form
\begin{align}
 \int_{-1}^{1}dz\,
 \frac{\sum_r c_{r,ab}^{(X,J)}(k_{0E},k_s;q_{0E},q_s)z^r}
      {\ell_E^2(z)+m_X^2} ,
 \label{eq:app-point-angular-integral}
\end{align}
apart from overall constants. Integration gives algebraic terms and the
logarithm in Eq.~\eqref{eq:angular-log-minus}. The Coulomb-gauge tensor
also contains $\boldsymbol\ell^2$ denominators and requires separate angular
moments.

For the monopole form factor in Eq.~\eqref{eq:CK-FF}, the boson denominator
and the two vertex form factors give the angular factor
\begin{align}
 \frac{(\Lambda^2-m_X^2)^2}
 {[\ell_E^2(z)+m_X^2][\ell_E^2(z)+\Lambda^2]^2}
 =\frac{1}{\ell_E^2(z)+m_X^2}
 -\frac{1}{\ell_E^2(z)+\Lambda^2}
 -\frac{\Lambda^2-m_X^2}{[\ell_E^2(z)+\Lambda^2]^2}.
 \label{eq:app-FF-decomposition}
\end{align}
Thus the regulated kernel contains logarithms with masses $m_X$ and
$\Lambda$ and algebraic terms. The massless singularities are treated as in
Sec.~\ref{subsec:method-1}.

\section*{Acknowledgements}
H. Q. Zhou acknowledges helpful discussions with Yong-Hui Lin, Zhi-Hui Guo, and Zhi-Yong Zhou. This work was funded by the National Natural Science Foundation of China (NSFC) under Grant No.~12675097.

\bibliography{ref}

\end{document}